# The Descent of Math

Sara Imari Walker
School of Earth and Space Exploration and Beyond Center for Fundamental Concepts in Science, Arizona State University, Tempe AZ USA
sara.i.walker@asu.edu

**Abstract:** A perplexing problem in understanding physical reality is why the universe seems comprehensible, and correspondingly why there should exist physical systems capable of comprehending it. In this essay I explore the possibility that rather than being an odd coincidence arising due to our strange position as passive (and even more strangely, conscious) observers in the cosmos, these two problems might be related and could be explainable in terms of fundamental physics. The perspective presented suggests a potential unified framework where, when taken together, comprehenders and comprehensibility are part of *causal structure* of physical reality, which is considered as a causal graph (network) connecting states that are physically possible. I argue that in some local regions, the most probable states are those that include physical systems which contain information encodings – such as mathematics, language and art – because these are the most highly connected to other possible states in this causal graph. Such physical systems include life and - of particular interest for the discussion of the place of math in physical reality – comprehenders capable of making mathematical sense of the world. Within this framework, the descent of math is an undirected outcome of the evolution of the universe, which will tend toward states that are increasingly connected to other possible states of the universe, a process greatly facilitated if some physical systems know the rules of the game. I therefore conclude that our ability to use mathematics to describe, and more importantly *manipulate*, the natural world may not be an anomaly or trick, but instead could provide clues to the underlying causal structure of physical reality.

Anthropic arguments are often used to explain away some of the most perplexing questions of our existence. Among these is the question of why the values of the constants of nature seem *surprisingly* well suited for life – indicative of a problematic degree of fine-tuning (dubbed the "fine-tuning problem"). However, one can reason, based on anthropic arguments, that the constants must be such as they are; otherwise we would not be here to ask about them [1]. We can similarly employ anthropic arguments to "solve" the problem of the rapidity of life's emergence, which happened almost as soon as conditions were favorable, and the related problem of an apparent arrow of increasingly biological complexity with time; both are necessary for us to be here and now asking how it could have happened [2]. While such appeals to anthropocentrism may not be entirely satisfactory to some, the reasoning is logically sound, challenging whether further explanation beyond anthropic arguments is indeed necessary.

There is however at least one anthropic feature of the universe as we observe it that cannot be adequately dismissed with these kinds of arguments. Namely, *why does the universe seem comprehensible and why should beings like us be here to comprehend it?* There appears to be no necessary reason why the universe should make sense to a subsystem of itself, and no necessary reason that we humans, being such a subsystem, should be able to make sense of it. Certainly we could exist without comprehending the world in a deep mathematical way (or for that matter making any sense of it at all (and indeed, one might argue that this is the *modus operandi* for most of us!)). While we would not be having an intellectual debate on the issue if we were not comprehenders, there is nothing about our existence that depends on our ability to make mathematical sense of the world in the same way that our existence is dependent on the fine-tuned value of the fine-structure constant (which if changed by just 4% would render formation of Carbon via stellar fusion impossible [3]). The situation is stranger still if we consider not only that we can make sense of the world by utilizing descriptive tools such as math, but that we can learn to do so as individuals, roughly on the timescale of a university education (see *e.g.,* [4] for discussion on this intriguing point).

It is easy enough to ignore these kinds of oddities, either as bizarre flukes or as pointless questions from the scientific standpoint. However, reality draws no dividing line between "artificial" and "natural" or, more to the point, "comprehender" and "everything else", as we have a tendency to do from our anthropocentric vantage point. Thus, the observational facts that the universe seems to make some sense to us, and that we can indeed (at least in part) make sense of it, must be taken into consideration in scientific discourse if we want to fully understand the universe in which we actually live (*e.g.,* one that includes comprehenders).

In example, a perfectly good scientific account exists for how Earth and other planetary bodies and natural satellites were formed via accretion from the solar nebulae, some 4.5 billion years ago. We do not have a similar explanatory framework that encompasses the "anti-accretion" happening in the Earth system today, where artificial satellites are being launched into orbit at an increasing pace [5]. The most important difference is that "anti-accretion" requires comprehenders – specifically, the existence of physical systems with *knowledge* of Newton's laws. To state this distinction more explicitly, the problem here does not arise in our descriptions of the orbital mechanics of the International Space Station (ISS) versus that of the Moon. Both would be much the same, with minor variation (*e.g.,* could be explained using Newtonian gravity, or more precisely general relativity). If instead, however, you were to ask how each was *caused* to enter its orbit, in the case of the Moon you could provide a perfectly good account based on the initial conditions of the Solar System and the laws of physics. In the case of the ISS, that account would necessarily also include *physical systems with knowledge of the laws of gravitation*.

Making this distinction about causation may seem unnecessary or even trivial, but doing so leads to a significantly different perspective on the structure of reality in the local vicinity of comprehenders, which I explicate in this essay. This perspective permits the incorporation of the existence of comprehensibility and comprehenders in a unified framework where, when taken together, comprehenders and comprehensibility are integral to the *causal structure* of physical reality. As such, rather than being an anomaly or bizarre fluke, the fact that we use information encodings – such as mathematics, language and even art – to describe and more importantly *manipulate* the world may in fact be a highly probable state in the space of all possible states of physical reality. It therefore could be explainable in terms of fundamental physics. In this formulation, the *"descent of math"* is an undirected product of the evolution of physical reality (under certain assumptions), just as the *"descent of man"* is an undirected product of the process biological evolution [6] (that is, neither requires design, although both might give the appearance of design (see *e.g.,* [7] for discussion on this matter)).

**Life and Physics: Two Roads Diverged in a Wood**

In physics we are trained to think in terms of initial conditions and deterministic, fixed laws of motion (the "prevailing conception" as discussed by Deutsch [8]). This has been an incredibly powerful approach to understanding systems as diverse as the interior of stars, superconductivity and swinging pendulums. However, at least thus far, this approach has fallen short of providing an adequate explanatory framework for life or mind. A challenge is that in describing life and related processes, we often use words such as "signaling", "symbols" and "codes" – that is, biology is cast in the language of *information*. The very concept of information, however, at present is not readily reconciled with a narrative cast solely in terms of initial conditions and fixed deterministic laws. A simple way to conceptualize this rift is to recognize that in order for something to carry information, it needs to have at least two possible states, *e.g.*, to say something carries 1 bit of information means that it could have been in one of two possible states. Fixed deterministic laws only allow one outcome at any given time $t$ for a given initial state at $t=0$, and therefore do not allow the possibility of more than one possibility. That is, they do not allow the possibility of *counterfactuals* – roughly described as situations that do not happen, but could. Our laws of physics permit a host of unrealized universes that could happen, but don't. Starting from a given initial condition, the future state of the universe as described by fixed dynamical laws is set for all time on one – *and only one* – of what could be many possible trajectories through state space.

We should contrast this with what we know of biology, which appears to be incredibly path-dependent. A good example is the path-dependence of biological evolution. Starting from the same initial state, biological systems trace out an enormous array of alternative trajectories through the process of evolution. Thus, a common statement in evolutionary biology such as "we share a common ancestor with $X$", where $X$ is a chimpanzee, fungi, or archaea, effectively is meant to tell us how far in the past we once shared a common initial state (to rough approximation, a genome) with species $X$. Each separate species evolved its unique features due to the peculiar and specific selection pressures it has seen through its evolutionary history, both due to the environment and competition with other organisms. Thus, in biology we are well acquainted with the fact that the current state is a function of (evolutionary) history. It is difficult, if not impossible, to write an equation of motion for such historical processes. The challenge is the *"state-dependent"* or *"self-referential"* nature of this kind of dynamical system [9]: the manner in which biological systems evolve through time is a function of their current state, such that the dynamical rules themselves change as a function of the current state [10]. The "laws" of biology therefore appear to be time dependent (this is a hallmark of self-referential systems and life, see *e.g.* [10, 12] for discussion). This naturally leads to historically dependent trajectories: the current state will depend on the sequence of previous states [11], masking dependence on initial conditions. State-dependent dynamics are deeply tied to the role of information in life: in part, the information encoded in the current state dictates what the system will do next.

What is intriguing about this situation is that biology appears to be taking advantage of precisely what is lacking in our ability to unify physics and information: biological systems are capable of taking many possible paths through state space, which must somehow be distinguished by their use of information. This observation has led many to suggest that the physics of biology *is* the physics of information, and in particular for my collaborator Paul Davies and I to suggest that the origin of life itself coincides with the emergence of physical systems where information plays a causal role (see *e.g.* [12] and references therein) perhaps representing an entirely new frontier in physics. Cast in the language of the discussion presented here, biological systems appear to represent a different kind of physics, one where information, in an as yet unspecified sense, seems necessary to define the trajectories taken through state space. To connect this story to the role of comprehenders, and our use of math to comprehend reality, we will need to take a detour and explore the structure of what is possible under the known laws of physics.

**You Can't Get There from Here**

Our world seems inordinately complex, being chock full of interesting things like bacteria, ant colonies, humans and cities. With our current formulation of physics, *all* of this complexity must be explained in terms of initial conditions and *fixed* dynamical laws. This leads to an odd accident as far as the status of initial conditions is concerned in the way we do physics. Since we do not seem to have the freedom to play with the laws, the forgoing indicates that any explanation we have for why the world is such as it is, and not any other way it could be, must be included in the specification of the initial state.

Take for example, a state of reality where Germany defeats Argentina to win the World Cup. Certainly an alternative state of the world – a counterfactual – might be Argentina defeating Germany. Neither scenario, describing Germany or Argentina emerging as the victor, violates the laws of physics, so it cannot be the laws themselves that distinguish which event happens[1]. This leaves open only the option that the initial condition is the distinguishing factor, *i.e.*, that the universe started in a very special initial state in which it was encoded that the Germany would win the World Cup on a planet called Earth, orbiting a humdrum star, in AD 2014. In the framework of initial conditions and fixed laws of motion as the sole descriptor of reality, the more complex the world becomes, the more special among the set of possible initial states must be the one that actually produced it (*i.e.*, the more fine-tuned it becomes). Apart from feeling unsatisfied with the idea that the universe's future for all time is laid down in its initial state (which indeed must necessarily be very special to explain our world), the focus we have in physics on initial states and fixed laws of motion is also somewhat at odds with a general feeling some of us may share that anything that is possible (allowed by the laws of physics) should be able to be caused to occur (in a laboratory for example).

---

[1] This may be a point of contention for some, as one might argue that Germany won due to physical superiority, but that is a different sort of physical argument than the one being put forth here.
[2] This constraint imposes all possible states of the world to have a one-to-one deterministic mapping connecting each state to at most two other states per time step.

This feeling has a bit of the flavor of statistical mechanics, where we base many of our calculations on the assumption of metric transitivity, which asserts that a system's trajectory through phase space will eventually explore the entirety of its state space – thus everything that is physically possible will eventually happen. That is, anything can be caused to occur by the laws themselves will happen, if only we wait long enough. While there are many examples of isolated systems that do not obey metric transitivity (possibly as a result of their underlying causal structure), the feature that is important for discussion here is that true metric transitivity makes initial conditions irrelevant, since every state will eventually be visited. In metrically transitive systems, "special" initial conditions that lead to restricted orbits through phase space, isolated from other possible trajectories, are excluded (are of measure zero).

A key point is that if we require specialness in our initial state (such that we observe the current state of the world and not any other state) metric transitivity is violated. It is then not necessarily possible to get to any other physically possible state – even those that may be equally consistent and permissible under the laws of physics. This leaves us in a bit of a perplexing situation, as we require special initial conditions to explain the complexity of the world, but also have a sense that we should not be on a particularly special trajectory to get here (or anywhere else) as it would be a sign of fine-tuning of the initial conditions. More simply put, a potential problem with the way we currently formulate physics is that you can't necessarily get everywhere from anywhere.

**Physical Reality and the Art of the Possible**

The key point of the above discussion is best clarified if we distinguish "possible" states of the world, defined as those that are not forbidden by the laws of physics, and "physically accessible" states of the world, which are those that are achievable from a given state (either in its "past" or "future"). Metric transitivity requires that all possible states are physically accessible, that is we assume systems where

$$\mathbb{N}_A \approx \mathbb{N}_P$$

where $\mathbb{N}_P$ are the number of physically possible states of the world, and $\mathbb{N}_A$ are those that can be realized (are physically accessible) from a given starting point.

However, a physics where $\mathbb{N}_A \approx \mathbb{N}_P$ has very strict constraints on its causal and informational structure and is not necessarily always metrically transitive. It is simplest to envision what these constraints are by considering *a causal graph (network) for reality*, where two instantaneous states of the universe are connected if, under the laws of physics, one state maps to the other. Edges connect nodes with their possible causes (states that map to the given node under the laws of physics) and effects (states that a given node maps to under the laws of physics). The constraint $\mathbb{N}_A \approx \mathbb{N}_P$ imposes a very restricted topology on this graph. For one, each possible state must have only one cause and one effect[2], and therefore each node has at most two edges. This constraint also means that logical microscopic reversibility holds and that no information is lost in the mapping between states (this is true because a cause or an effect are fully specified by looking at the current state, since the mapping is one-to-one). In such systems, the trajectories through state space follow fixed paths along closed causal loops. In many dynamical systems, such as cellular automata, these causal loops are disconnected, so while $\mathbb{N}_A \sim \mathbb{N}_P$, metric transitivity does not hold because there are isolated parts of the graph. This is probably true in the real world, but we often do not consider the underlying causal structure in our theories, particularly in statistical physics where it is assumed that any state can transition to any other with a typically small but nonetheless finite probability (that is, we ignore causal structure).

By contrast situations where $\mathbb{N}_A \not\approx \mathbb{N}_P$ arise when the causal structure is such that multiple causes map to one effect (or a cause has multiple effects). Nodes can have any number of edges (up to a number connecting a node to any possible state). An important point about this causal architecture for physical reality is that information is lost in mapping between states for a many-to-one map: if you ran the system in reverse, there would be uncertainty in which state was the initial cause. This has potentially deep connections to information loss and the emergence of an arrow of time, which is normally associated with information loss due to coarse-graining [13]. In this formalism, states are connected when they share information[3]. Importantly, the number of states accessible from a node (its effects) is precisely its number of its out-directed edges in the network graph of reality. Likewise, the number of states a given node can be accessed from (its causes) is the number of in-directed edges.

---

[2] This constraint imposes all possible states of the world to have a one-to-one deterministic mapping connecting each state to at most two other states per time step.
[3] This last point is most obvious for the case of one-to-one mapping where no information about past states is lost in the mapping.

An example of why any of this should matter may be in order, so we will return to our initial discussion of Earth and its satellites. Consider the set of all states of reality that correspond to Earth plus satellites. This set may be partitioned into two subsets: states where some physical systems have knowledge of Newton's laws[4] (*i.e.,* comprehenders, such as humans) and states where there is no such knowledge instantiated in physical systems. One example of the latter is Earth with one natural satellite – the Moon, which was the state of Earth for its 4.5 billion year history prior to 1957. It is physically possible that Earth might have also had no Moon, two captured asteroids for Moons (as is the case for Mars) or a potential host of alternative smallish rocky bodies orbiting it. All are equally viable states of the world consistent with known physics. It is also physically possible for the Earth have any number (within resource constraints) of artificial satellites or space junk. However, the latter set of "anti-accreting" states, while possible, is not accessible (by this I mean *not* encoded in the initial conditions) in the absence of comprehenders and the technology they create. Thus, when comparing the size of the state space of Earth plus satellites, the state space is *much larger* if artificial satellites and the comprehenders that launch them to space are included, *i.e.,* if their exist physical systems that contain knowledge of the laws of gravitation (as well as good engineers).

The above suggest there is in fact another kind of edge in the causal graph of reality in addition to those set by the laws of physics – that is, some edges are set by physical systems that contain knowledge of the laws of physics. In an *"explanatory graph of reality"*, some of the causal edges between states are set by the existence of physical systems that contain knowledge about other possible states. Here, 'knowledge' need not be anything nearly as sophisticated as the law of universal gravitation, but can include simpler rules encoded in physical systems, such as that swimming up a chemical gradient will likely result in finding a new food source, as is the case for bacterial chemotaxis. Knowledge in the sense presented here is therefore directly related to the number of counterfactuals about *what could be caused to happen* instantiated in a given physical system. This approach therefore has some promise for quantifying the murky concept of knowledge in physics precisely by identifying the knowledge in a physical system with the number of out-directed edges in the explanatory graph (or conversely, as the number of counterfactuals instantiated in the physical system). Physical instantiation of knowledge of *theories* like Newton's laws is particularly powerful since it connects a large number of states that would otherwise be disconnected (hence it is corresponds to a large number of "explanatory" edges or knowledge and is made a "law").

---

[4] To be perfectly accurate I should include general relativity in this argument, but the argument stands the same regardless of what our current theories to describe reality are, so long as they permit new states of the world to be realized that could not be realized in the absence of such knowledge.

There is a second important reason why knowledge of Newton's law manifests differently in the explanatory graph than the edges set by Newton's law itself. To illustrate this, we return again to our two aforementioned examples of the Earth with natural and artificial satellites. In the case of the Earth and Moon, described under a scenario of initial conditions and deterministic laws, only the states preceding the giant Moon forming impact could have "caused" the Moon to form. A second ISS, by contrast, could easily be launched to space within a few months if there was sufficient will power to do so. Comprehenders therefore have the property that they can more reliably cause a transformation to occur (in the sense that it can happen again) than the laws of physics alone can. This notion of reliability is an important concept in constructor theory [7, 8], where "constructors" are identified with reliable causes with the important property that constructors performing a transformation retain the ability to do so again (are reliable).

The combination of increased connectivity and reliability in the vicinity of comprehenders or knowledge-creating systems provides a potential explanation for the situation we find ourselves in as comprehenders occupying a seemingly comprehensible reality. When considering the number of states of the world consisting of Earth plus satellites, very many more states are reliably accessed from states in which some physical systems have knowledge of the laws of gravitation than from those where there is no such information. Thus, in a network view of reality, where causal edges connect all physically possible states, nodes with physical systems (comprehenders) that contain knowledge of Newton's law are more highly connected to other nodes within the space of all configurations of Earth plus satellites.

Reliability implies that states that arise as a result of traversing a trajectory connected by knowledge should, on average, retain the knowledge of past states along the trajectory (such that transformations that were possible in the past remain possible). Thus, connectivity should be "heritable" among nodes with knowledge. Such *heritable connectivity* could provide a physical explanation for the ubiquity of reproduction in biological systems, reproduction is a surefire method of ensuring heritability of information encodings necessary to instantiate knowledge (retain causal edges).

The essential point of this argument in connecting the existence of comprehenders and comprehensibility is that states of reality containing physical systems with knowledge (comprehenders) should be more *probable* than states that don't, once one accounts for the underlying causal structure. What I mean by probable does not strictly refer to counting the frequency of a state (as is traditionally done in statistical physics), which does not account for the existence of any underlying causal structure. The most probable states here are those that are the most highly connected via knowledge (have the largest number of edges) in the causal graph because they are the most likely to be visited, if one assumes a random walk along the graph with the principle of reliable causes. I argue that those most probable states should include comprehenders, because through knowledge, comprehenders connect many states of reality that would otherwise be unconnected and do so more reliably than the laws of physics (connectivity is heritable). These states must also be comprehensible, as otherwise knowledge would not result in connections to those states. Both comprehenders and comprehensibility are therefore required to create causal edges (in other words, the two states must share information).

**Life, Information and … Math!**

We should not necessarily immediately assume that reality is structured such that all states are accessible, or reliably accessible (and in fact as noted above, for reversible dynamical systems with $\mathbb{N}_A \approx \mathbb{N}_P$ all states are accessible from somewhere in the causal graph, but in most cases I think we would find none are from everywhere). *What kind of scenario then would allow the most states within a local region of the causal graph of reality to be realizable?* I suggest this requires physical systems that actively use information to move through state space; in short it requires life, mind and related processes. The impact on the world when we discover a new physical law – *e.g.*, a pattern in how the world works – behaves much more like the process of biological evolution, than (perhaps ironically) it does like the physical systems we describe with those laws. Biological evolution accesses many states of the world from similar starting conditions. Likewise, the discovery of new laws of nature allows many states of the world to potentially be accessed that were not accessible before their discovery. At their core, both of these processes have one important thing in common: multiple states of the world can be caused to occur, which are not explicitly encoded in the initial condition (both necessitate states which are highly connected).

The physics necessary to describe such processes requires more than just an initial condition and deterministic law – *it requires information*. In this formalism, knowledge about reality allows physical systems to access states of the world not encoded in their initial conditions, by making more than one state accessible. Physical systems, which encode information about other states of the world (connected by an edge) and where that information in part defines their trajectory through state space, will traverse non-trivial dynamical trajectories that are history and path-dependent. Although I have been discussing this dynamic as though time is discrete, discrete time is not central to the argument.

In the framework presented, the most probable among all possible states of physical reality are those that include physical systems which contain information encodings – such as mathematics, language and art – because these are the most highly connected states in the local state space of what is possible. That is to say, they have the highest number of reliable edges (satisfying heritable connectivity) in the network of the possible states of the world. A random walk on the network of all that is possible will spend the most time on highly connected nodes, and is most likely to follow a trajectory where the connectivity of nodes increases as a function of time.

This suggests a generalization of the second law of thermodynamics to the scenario where $\mathbb{N}_A \not\approx \mathbb{N}_P$, where instead of the most commonly visited states being those which are most probable, it is instead those *that are most highly connected*. This formalism naturally accommodates an arrow of increasing complexity (knowledge), corresponding to an arrow pointing towards states that have increasing information about other possible states of the world [12]. Such an arrow of knowledge would likely have many of the hallmarks associated with the arrow of complexity in the biosphere. Within this framework, the descent of math is an undirected, but not unexpected, outcome of the evolution of the universe, which will tend toward states that are increasingly connected to other possible states of the universe, a process greatly facilitated if some physical systems know the rules of the game. This framework is still under development, but it shows promise as both *comprehenders* – defined as physical systems where instantiated information about the world (knowledge) in part determines their trajectory through state space, *and comprehensibility* – defined in terms of states that can share information (edges created by knowledge), are naturally accommodated.  Our ability to use mathematics to describe, and more importantly *manipulate*, the natural world may not be an anomaly or "trick", but instead could provide important clues to the causal structure of physical reality.

**Acknowledgements**
The author wishes to thank Paul Davies and Chiara Marletto for comments, and in particular to thank Chiara Marletto for the terminology of "explanatory graph" to describe the network of reality containing comprehenders and the insight that heritable connectivity is essential to its formulation. Additional thanks to the FQXi community for the lively discussion and insightful comments on this essay